\PassOptionsToPackage{x11names,table,rgb,html}{xcolor}
\documentclass[british,screen,nonacm]{acmart}
\pdfoutput=1

\usepackage{xparse}
\usepackage{xpatch}
\usepackage{xspace}
\usepackage{etoolbox}
\usepackage{environ}

\usepackage[T1]{fontenc}
\usepackage[utf8]{inputenc}
\usepackage{microtype}

\usepackage{xcolor}

\usepackage[british]{babel}
\usepackage[all]{foreign}
\defnotforeign[wrt]{{\xperiodafter{w.r.t}}}
\defnotforeign[wl]{{\xperiodafter{w.l.o.g}}}

\usepackage{amsmath}
\usepackage{amsfonts}
\usepackage{amssymb}
\usepackage{stmaryrd}
\usepackage{latexsym}
\usepackage{mathtools}
\usepackage{esvect}
\renewcommand{\vec}{\vv}

\usepackage{bussproofs}

\SetSymbolFont{stmry}{bold}{U}{stmry}{m}{n}

\allowdisplaybreaks

\def\cspjot{7pt}

\AtEndPreamble{
	\theoremstyle{acmdefinition}

}

\usepackage{float}
\usepackage{wrapfig}
\usepackage{placeins}

\usepackage{booktabs}

\usepackage{csvsimple}
\makeatletter
\csvset{
  autobooktabularcenter/.style={
    file=#1,
    before reading=\small, 
    after head=\csv@pretable\rowcolors{2}{gray!10}{white}\begin{tabular}{*{\csv@columncount}{c}}\csv@tablehead,
    table head=\toprule\csvlinetotablerow\\\midrule,
    late after line=\\,
    table foot=\\\bottomrule,
    late after last line=\csv@tablefoot\end{tabular}\csv@posttable,
    command=\csvlinetotablerow},
}
\makeatother
\newcommand{\csvautobooktabularcenter}[2][]{\csvloop{autobooktabularcenter={#2},#1}}

\usepackage[inline]{enumitem}

\usepackage{graphicx}
\usepackage{subcaption}
\usepackage{tikz}
\usetikzlibrary{calc,fit,shapes,positioning,arrows,intersections}

\usepackage{hyperref}
\usepackage{nameref}

\usepackage[capitalise,nameinlink,noabbrev]{cleveref}

\makeatletter
\newcommand{\customlabel}[4][0]{%
	\protected@write\@auxout{}{\string\newlabel{#3}{{#4}{\thepage}{#4}{#3}{}}}%
	\protected@write\@auxout{}{\string\newlabel{#3@cref}{{[#2][#1][#1]#4}{\thepage}}}%
}
\newcommand{\crefv}[1]{%
	\begingroup\@cref@compressfalse\@cref@sortfalse\cref{#1}\endgroup%
}
\newcommand{\Crefv}[1]{%
	\begingroup\@cref@compressfalse\@cref@sortfalse\Cref{#1}\endgroup%
}
\makeatother

\hypersetup{
	hidelinks,
	final,
}

\setcitestyle{numbers,sort&compress}
\usepackage[disable]{todonotes}

\setcopyright{none}

\makeatletter
\let\orgdescriptionlabel\descriptionlabel
\renewcommand*{\descriptionlabel}[1]{%
  \let\orglabel\label
  \let\label\@gobble
  \phantomsection
  \edef\@currentlabel{\ignorespaces #1\unskip}%
  \let\label\orglabel
  \orgdescriptionlabel{#1}%
}
\makeatother

\newcommand{\N}{\mathbb{N}}
\newcommand{\defeq}{\triangleq}

\newcommand{\mono}{\rightarrowtail}

\newcommand{\emb}{\hookrightarrow}
\newcommand{\face}[1]{\langle #1 \rangle}

\newcommand{\rng}{\textsf{\textit{img}}}
\newcommand{\prnt}{\textsf{\textit{prnt}}}
\newcommand{\ctrl}{\textsf{\textit{ctrl}}}
\newcommand{\link}{\textsf{\textit{link}}}
\newcommand{\ephi}[1]{\phi^\mathsf{#1}}
\newcommand\restr[2]{{\left.\kern-\nulldelimiterspace#1\vphantom{\big|}\right|_{#2}}}
\newcommand{\cat}[1]{\textsc{#1}}

\newenvironment{myfont}{\fontfamily{qhv}\selectfont}{\par}
\DeclareTextFontCommand{\chfont}{\myfont}

\def\libbig{\textsf{\href{http://mads.uniud.it/downloads/libbig/}{jLibBig}}\xspace}

\begin{document}

\title{A CSP implementation of the directed bigraph embedding problem}

\author{Alessio Chiapperini}
\author{Marino Miculan}
\orcid{0000-0002-7866-7484}
\affiliation{
  \institution{University of Udine}
  \streetaddress{Via delle Scienze 206}
  \city{Udine}
  \postcode{33100}
  \country{Italy} 
}
\email{chiapperini.alessio@spes.uniud.it}
\email{marino.miculan@uniud.it}

\author{Marco Peressotti}
\orcid{0000-0002-0243-0480}
\affiliation{
  \institution{University of Southern Denmark}
  \streetaddress{Campusvej 55}
  \city{Odense}
  \postcode{5230}
  \country{Denmark} 
}
\email{peressotti@imada.sdu.dk}

\begin{abstract}
\emph{Directed bigraphs} are a meta-model which generalises Milner's bigraphs by taking into account  the \emph{request flow} between controls and names.
A key problem about these bigraphs is that of \emph{bigraph embedding}, i.e., finding the embeddings of  a bigraph inside a larger one. 
We present an algorithm for computing embeddings of directed bigraphs, via a reduction to a \emph{constraint satisfaction problem}.
We prove soundness and completeness of this algorithm, and provide an implementation in \libbig, a general Java  library for manipulating bigraphical reactive systems, together with some experimental results.
\end{abstract}



\maketitle

\section{Introduction}

\emph{Bigraphical Reactive Systems} (BRSs) are a family of graph-based formalisms introduced as a meta-model for distributed, mobile systems \cite{jensen2003bigraphs,milner2009space,miculan2013presheaves}. 
In this approach, system configurations are represented by \emph{bigraphs}, graph-like data structures capable of describing at once both the locations and the logical connections of (possibly nested) components.  
The dynamics of a system is defined by means of a set of \emph{graph rewriting rules}, which can replace and change components' positions and connections.
BRSs have been successfully applied to the formalization of a wide spectrum of domain-specific models, including context-aware systems, web-service orchestration languages  \cite{bundgaard2008formalizing,mansutti2014multi,sahli2019self,burco2019towards}.
BRSs are appealing because they provide a range of general results and tools, which can be readily instantiated with the specific model under scrutiny: libraries for bigraph manipulation (e.g., DBtk \cite{bacci2009dbtk} and \libbig\ \cite{miculan2014csp,jlibbig}), simulation tools \cite{mansutti2015distributed,mansutti2014towards,gassara2019executing}, graphical editors \cite{faithfull2013big}, model checkers \cite{perrone2012model}, modular composition \cite{perrone2011bigraphical}, stochastic extensions \cite{krivine2008stochastic}, etc.

\begin{figure}[t]
	\centering
  \includegraphics[width=.7\textwidth]{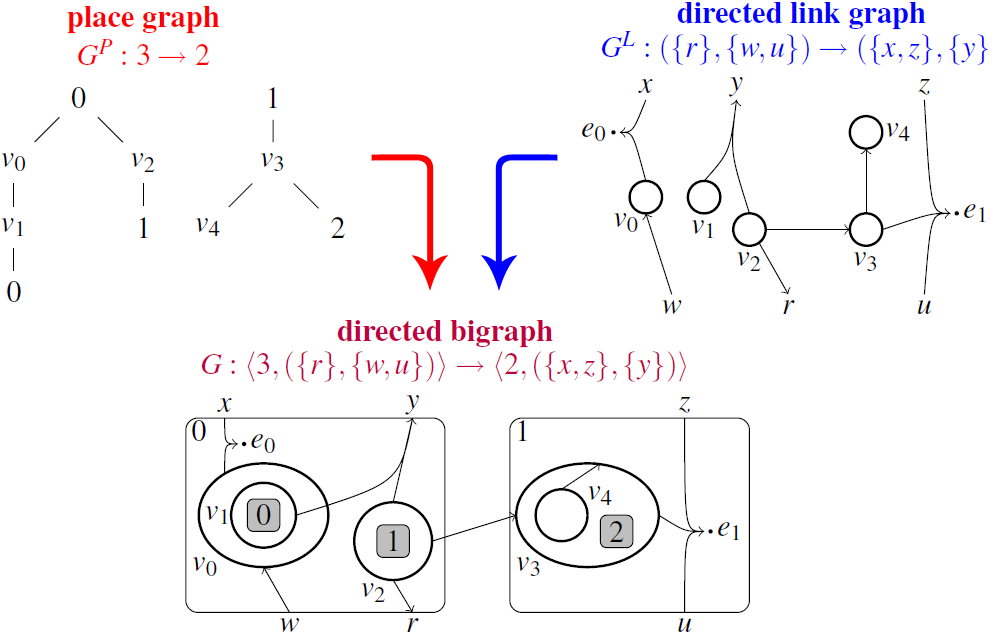}
	\caption{An example of directed bigraph and its place and link graphs \cite{grohmann2008access}.}
	\label{fig:example1}
\end{figure}

\looseness=-1
Along this line, \cite{grohmann2007directed,gm:concur07} introduced \emph{directed bigraphs}, a strict generalization of Milner's bigraphs where the link graph is directed (see \cref{fig:example1}).
This variant is very suited for reasoning about \emph{dependencies} and \emph{request flows} between components, such as those found in producer-consumer scenarios.
In fact, they have been used to design formal models of 
security protocols \cite{grohmann2008security},
molecular biology \cite{bacci2009framework},
access control governance \cite{grohmann2008access},
container-based systems \cite{burco2019towards}, among others.

A key notion at the core of these results and tools is that of \emph{bigraph embedding}.
Informally, an embedding is a structure preserving map from a bigraph (called \emph{guest}) to another one (called \emph{host}), akin a subgraph isomorphism. Finding such embeddings is a difficult problem; in fact, the sole embedding of place graphs has been proved to be NP-complete \cite{bacci2014finding}. 
Several algorithms have been proposed in literature for bigraphs with undirected links (see e.g. \cite{glenstrup2007implementation,sevegnani2010sat,damgaard2013inductive,miculan2014csp}), but there are no embedding algorithms for the more general case of directed bigraphs, yet.

In this work, we propose an algorithm for computing embedding of directed bigraphs (which subsume traditional ones), laying the theoretical and technical foundations for bringing directed bigraphs to tools like the ones listed above.

More precisely, in \cref{sec:brs} we first introduce directed bigraphs and bigraphic reactive systems, generalizing \cite{grohmann2007directed,burco2019towards}.
Then, the notion of directed bigraph embedding is defined in \cref{sec:emb}.
In \cref{sec:csp} we present a reduction of the embedding problem for directed bigraphs to a constraint satisfaction problem (CSP) and show that it provides a sound and complete algorithm for computing embeddings. This reduction extends our previous (unpublished) work  \cite{miculan2014csp} on the embedding problem for undirected bigraph.
We have implemented this algorithm as an extension of \libbig \cite{jlibbig}, a general Java library for BRSs; this implementation and some experimental results are reported in \cref{sec:experiments}. 
Finally, some conclusions and directions for future work are drawn in \cref{sec:concl}.

\section{Reactive systems on Directed Bigraphs}
\label{sec:brs}
 
In this section we introduce \emph{directed} link graphs and bigraphs, and \emph{directed bigraphical reactive systems}, extending the definitions from \cite{grohmann2007directed,gm:concur07}.

\subsection{Directed bigraphs}
\begin{definition}[Polarized interface]
	A \emph{polarized interface} $X$ is a pair $(X^-, X^+)$, where $X^+$ and $X^-$ are sets of names s.t. $X^- \cap X^+ = \varnothing$; the two sets are called \emph{downward} and \emph{upward} interfaces respectively.
\end{definition}

\begin{definition}[Polarized signature]
	A \emph{signature} is a pair $(\mathcal{K}, ar)$, where $\mathcal{K}$ is the set of \emph{controls}, and $ar :	\mathcal{K} \to \N \times \N$ is a map assigning to each control its polarized arity, that is, a pair $ \langle n, m \rangle$ where $n, m$ are the numbers of \emph{positive} and \emph{negative} ports of the control, respectively.
	
	We define $ar^+, ar^- : \mathcal{K} \to \N$ as shorthand for the \emph{positive} and \emph{negative} ports of controls: $ar^+ \defeq \pi_1 \circ ar$, $ar^- \defeq \pi_2 \circ ar$.
\end{definition}

The main difference between this definition and that from~\cite{grohmann2007directed} is that we allow also for \emph{inward} ports in controls, whereas  in \cite{grohmann2007directed}, like in \cite{milner2009space}, controls have only outward ports.
This turns out also in the definition of \emph{points} and \emph{handles}.

\begin{definition}[Directed Link Graph]
	A directed link graph $A: X \to Y$ is a quadruple $A = (V,E,ctrl,link)$ where $X,Y,V,E$ and $ctrl$ are
	defined as before, while the link map is defined as $link: Pnt(A) \to Lnk(A)$ where
	\begin{alignat*}{3}
		Prt^+(A) \triangleq {} & \sum_{v \in V} ar^+(ctrl(v)) 
    &\quad
		Prt^-(A) \triangleq {} & \sum_{v \in V} ar^-(ctrl(v))
    \\
		Pnt(A) \triangleq {} & X^+ \uplus Y^- \uplus Prt^+(A) 
    &
    Lnk(A) \triangleq {} & X^- \uplus Y^+ \uplus E \uplus Prt^-(A)
  \end{alignat*}
  with the following additional constraints:
  \begin{gather}
      \forall x \in X^-,\forall y \in X^+ . link(y) = x \Rightarrow link^{-1}(x) = \{y\}  \label{eq:innerconst}\\
      \forall y \in Y^+,\forall x \in Y^- . link(x) = y \Rightarrow link^{-1}(y) = \{x\}.  \label{eq:outerconst}
  \end{gather}	
	The elements of $Pnt(A)$ are called the \emph{points} of $A$; the elements of $Lnk(A)$ are called the \emph{handles} of $A$.	
\end{definition}
The constraint (\ref{eq:innerconst}) means that if there is an upward inner name connected to a downward inner name, then nothing else can be connected to the latter; constraint  (\ref{eq:outerconst}) is similar, on the outer interface. Together, these requirements guarantee that composition of link graphs (along the correct interfaces) is well defined.

Direct link graphs are graphically depicted similarly to ordinary link graphs, with the
difference that edges are represented as vertices of the graph and not as hyper-arcs connecting points and names.

Directed bigraphs are composed by a directed link graph and a place graph. 
Since the definition of place graph is the same as for pure bigraphs, we
will omit it and refer the interested reader to~\cite{milner2009space}.

\begin{definition}[Directed Bigraph]
	An \emph{interface} $I=\langle m,X\rangle$ is composed by a finite ordinal $m$, called the \emph{width}, and by a directed interface $X=(X^-,X^+)$. 

	Let $I = \langle m, X \rangle$ and $O = \langle n, Y \rangle$ be two interfaces; a \emph{directed bigraph} with signature	$\mathcal{K}$ from $I$ to $O$ is a tuple $G = (V,E,ctrl,prnt,link): I \to O$ where
	\begin{itemize}
		\item $I$ and $O$ are the \emph{inner} and \emph{outer} interfaces;
		\item $V$ and $E$ are the sets of nodes and edges;
		\item $ctrl,prnt,link$ are the \emph{control}, \emph{parent} and \emph{link} maps;
	\end{itemize}
	such that 
	$G^L \triangleq (V,E,ctrl,link): X \to Y$ is a directed link graph and
	$G^P \triangleq (V,ctrl,prnt): m \to n$ is a place graph, that is, the map $prnt: m\uplus V \to n\uplus V$ is acyclic.
	The bigraph $G$ is denoted also as $\langle G^P, G^L \rangle$. 
\end{definition}

Directed bigraphs can be composed on matching interfaces.
\begin{definition}[Composition and identities]\label{def:dir_comp}
	The composition of two place graphs $F : k \to m$ and $G : m \to n$, is defined in the same way as pure bigraphs (i.e., suitable grafting of forests).
  If $F : X \to Y$ and $G : Y \to Z$ are two link graphs,
	their composition is the link graph
    $G \circ F \defeq (V, E, ctrl, link) : X \to Z$
	such that $V = V_F \uplus V_G$, $E = E_F \uplus E_G$, $ctrl = ctrl_F \uplus ctrl_G$, and 
			$link : Pnt(G \circ F) \to Lnk(G \circ F)$ is defined as follows:
			\begin{align*}
					Pnt(G \circ F) & = X^+ \uplus Z^- \uplus Prt^+(F) \uplus Prt^+(G) \\
Lnk(G \circ F) & = X^- \uplus Z^+ \uplus Prt^-(F) \uplus Prt^-(G) \uplus E \\
				link(p) & \defeq
				\begin{cases*}
					prelink(p) & if $prelink(p) \in Lnk(G \circ F)$\\
					link(prelink(p)) & otherwise
				\end{cases*} 
			\end{align*}
		where $prelink : Pnt(G \circ F) \uplus Y^+ \uplus Y^- \to Lnk(G \circ F) \uplus Y^+$ is  $link_F \uplus link_G$.
		
			The identity link graph at $X$ is $\chfont{id}_X \defeq (\varnothing,
			\varnothing, \varnothing_{\mathcal{K}}, \chfont{Id}_{X^- \uplus X^+}) : X \to X$.

	If $F : I \to J$ and $G : J \to K$ are two bigraphs, their composite is $G \circ F \defeq \langle G^P \circ F^P, G^L \circ F^L \rangle: I \to K$
			and the identity bigraph at $I = \langle m, X \rangle$ is $\langle \chfont{id}_m, \chfont{id}_{X^- \uplus
			X^+} \rangle$.
\end{definition}

\newpage
\begin{definition}[Juxtaposition]
  For place graphs, the juxtaposition of two interfaces $m_0$ and $m_1$ is $m_0 + m_1$; the unit is 0.
			If $F_i = (V_i, ctrl_i, prnt_i) : m_i \to n_i$ are disjoint place graphs (with $i = 0,1$), their juxtaposition
			is defined  as for pure bigraphs.
  For link graphs, the juxtaposition of two (directed) link graph interfaces $X_0$ and $X_1$ is 
			$(X_0^- \uplus X_1^-, X_0^+ \uplus X_1^+)$. If $F_i = (V_i, E_i, ctrl_i, link_i) : X_i \to Y_i$ 
			are two link graphs (with $i = 0,1$), their juxtaposition is
			\begin{equation*}
				F_0 \otimes F_1 \defeq (V_0 \uplus V_1, E_0 \uplus E_1, ctrl_0 \uplus ctrl_1, link_0 
				\uplus link_1) : X_0 \otimes X_1 \to Y_0 \otimes Y_1
			\end{equation*}
  For bigraphs, the juxtaposition of two  interfaces $I_i = \langle m_i, X_i \rangle$ (with $i =
			0,1$) is $\langle m_0 + m_1, (X_0^- \uplus X_1^-, X_0^+ \uplus X_1^+) \rangle$ (the unit is 
			$\epsilon = \langle 0, (\varnothing, \varnothing) \rangle$). If $F_i : I_i \to J_i$ are two bigraphs 
			(with $i = 0,1$), their juxtaposition is
			\begin{equation*}
				F_0 \otimes F_1 \defeq \langle F_0^P \otimes F_1^P, F_0^L \otimes F_1^L \rangle : I_0
				\otimes I_1 \to J_0 \otimes J_1.
			\end{equation*}
\end{definition}

Polarized interfaces and directed bigraphs over a given signature $\mathcal{K}$ form a monoidal category $\cat{DBig}(\mathcal{K})$. 
It is worthwhile noticing that Milner's pure bigraphs \cite{milner2009space} correspond precisely to directed bigraphs between ascending (i.e., positive) interfaces only and over signatures with only positive ports.

\subsection{Reactive systems over directed bigraphs}
In order to define reactive systems over bigraphs, we need to define how a parametric reaction rule (i.e., a pair of ``redex-reactum'' bigraphs) can be instantiated.
Essentially, in the application of the rule, the ``sites'' of the reactum must be filled with the parameters appearing in the redex. This relation is expressed by an \emph{instantiation map} in the rule.

\begin{definition}[Instantiation map]
    An \emph{instantiation map} $\eta :: \langle m,X \rangle \to \langle m',X' \rangle$ is a pair $\eta = (\eta^P, \eta^L)$ where 
    \begin{itemize}
    \item	$\eta^P :m' \to m$  is a function which maps sites of the reactum to sites of the redex; for each	$j \in m'$, it determines that the $j$-th site of the reactum is filled with the $\eta(j)$-th parameter of the redex.
    	
     \item $\eta^L : \left( \sum_{i=0}^{m'-1} X \right) \to X'$ is a wiring (i.e., a link graph without nodes nor edges), which is responsible for mapping names of the redex to names of the reactum. 
     This can be described as a pair of functions $\eta^L= (\eta^+, \eta^-)$  where $\eta^+ : \left( \sum_{i=0}^{m'-1} X^+ \right) \to X'^+$ and
    	$\eta^- : X'^- \to \sum_{j=0}^{m'-1} X^-$.
    \end{itemize}
\end{definition}

The dynamics of bigraphs is expressed in terms of rewriting rules.
\begin{definition}[Parametric reaction rule]
	A \emph{parametric reaction rule} for bigraphs is a triple of the form
  $		(R:I \to J, R':I' \to J, \eta : I \to I')$
	where $R$ is the parametric redex, $R'$ the parametric reactum and $\eta$ is an instantiation map.
\end{definition}

We can now define the key notion of reactive systems over directed bigraphs, which is a generalization of that in \cite{milner2009space,gm:concur07}.
Let $Ag(\mathcal{K})$ be the set of \emph{agents} (i.e., bigraphs with no  inner names nor sites) over a signature $\mathcal{K}$. 
\begin{definition}[DBRS]\label{def:dbrs}
    A \emph{directed bigraphical reactive system} (DBRS) $DBG(\mathcal{K}, \mathcal{R})$ is defined by a signature  $\mathcal{K}$ and a set $\mathcal{R}$ of rewriting rules.
    A DBRS $DBG(\mathcal{K}, \mathcal{R})$ induces a rewriting relation ${\rightarrowtriangle} \subseteq Ag(\mathcal{K}) \times Ag(\mathcal{K})$  via the derivation rule:
    \[
     \begin{array}{c}
     (R_L, R_R, \eta) \in \mathcal{R} \\
	 A = C \circ (R_L \otimes Id_Z) \circ \omega \circ (D_0 \otimes \ldots \otimes D_{m-1}) \qquad 
	\textstyle A' = C \circ (R_R \otimes Id_Z) \circ \omega' \circ (D_{\eta^P(0)} \otimes \ldots \otimes D_{\eta^P(m'-1)}) \\\hline
       A \rightarrowtriangle A'
    \end{array}
    \]
    where the \emph{wiring maps} $\omega$ and $\omega'$ are give as follows:
    \begin{alignat*}{3}
        \omega & : \sum_{i=0}^{m-1} X_i \to X \oplus Z 
        &
        \omega' &: \sum_{j=0}^{m'-1} X_{\eta^P(j)} \to X' \oplus Z 
        \\
        \omega^+ & : \sum_{i=0}^{m-1} X_i^+ \to X^+ \uplus Z^+
        &
        \omega'^+ &: \sum_{j=0}^{m'-1} X^+_{\eta^P(j)} \to X'^+ \uplus Z^+ 
        \\
        \omega^- &: X^- \uplus Z^- \to \sum_{i=0}^{m-1} X_i^-  \qquad
        &
        \omega'^- &: X'^- \uplus Z^- \to \sum_{j=0}^{m'-1} X^-_{\eta^P(j)} 
     \end{alignat*}
     \begin{align*}
        \omega'^+(j,x) & \triangleq
    			\begin{cases*}
    				\eta^+(j, \omega^+(\eta(j),x)) & if $\omega^+(\eta^P(j),x) \in X^+$\\
    				\omega^+(\eta^P(j),x) & if $\omega^+(\eta(j),x) \in Z^+$
    			\end{cases*}
    		\\
    		\omega'^-(x) & \triangleq (j,y) \text{ for } j \in \eta^{P^{-1}}(i) \text { and } (i,y) \in \eta^-(x)
     \end{align*}
\end{definition}

The difference with respect to the previous versions of BRS is that now links can descend from the redex (and reactum) into the parameters, as it is evident from the fact that redexes and reactums in rules may have generic inner interfaces ($I$ and $I'$). This is very useful for representing a request flow which goes ``downwards'', e.g. connecting a port of a control in the redex to a port of an inner component (think of, e.g., a linked library). 

However, this poses some issues when the rules are not linear.
If any of $D_i$'s is cancelled by the rewriting, the controls in it disappear as well, and we may be not able to connect some name descending from $R_L$ or $Id_Z$ anymore.
More formally, this means that the map $\omega^-$ can be defined only if for every $x \in (X'^- \uplus Z^-)$ there are $j,y$ such that $(\eta^P(j),y) = \eta^-(x)$. We can have two  cases:
\begin{enumerate}
    \item for some $x$, there are no such $j,y$. This means that $\omega$ is not defined and hence the rule cannot be applied.
    \item for each $x$, there are one or more pairs $(j,y)$ such that $(\eta^P(j),y) = \eta^-(x)$.
    This means that for the given source agent decomposition, there can be several ways to define $\omega^-$. Each of these possible definitions yields a different application of the same rule (and the same decomposition).
\end{enumerate}
Overall, the presence of downward names in parameters adds a new degree of non-determinism to Directed BRSs, with respect to previous versions of BRSs.

\section{Directed Bigraph embeddings}\label{sec:emb}
As we have seen in the previous section, to execute or simulate a BRS it is necessary to solve the \emph{bigraph matching problem}, that is, finding the occurrences of a redex $R$ within a given bigraph $A$.
More formally, this translates to finding $C, Z, \omega$ and $D = (D_0 \otimes \ldots \otimes D_{m-1})$ such that $A = C \circ (R\otimes Id_Z) \circ\omega \circ D$.
$C$ and $D$ are called \emph{context} and \emph{parameter}. 

If we abstract from the decomposition of the agent $A$ in context, redex and parameter we can see how the matching problem is related to the \emph{subgraph isomorphism} problem.
Therefore, in this Section we define the notions of \emph{directed bigraph embedding}. The following definitions are taken
from~\cite{hojsgaard2012bigraphical}, modified to suit the definition of directed bigraphs.

\paragraph{Directed Link graph}\looseness=-1
Intuitively an embedding of link graphs is a structure preserving map from one link graph (the \emph{guest}) to another
(the \emph{host}).  As one would expect from a graph embedding, this map contains a pair of injections: one for the
nodes and one for the edges (i.e., a support translation). The remaining of the embedding map specifies how names of
the inner and outer interfaces should be mapped into the host link graph.  Outer names can be mapped to any link; here
injectivity is not required since a context can alias outer names. Dually, inner names can be mapped to hyper-edges linking sets of points in the host link graph and such that every point is contained in at most one of these sets.

\begin{definition}[Directed link graph embedding]\label{def:dlge}
	Let $G:X_G \to Y_G$ and $H:X_H \to Y_H$ be two directed link graphs. A directed link graph embedding
	$\phi : G \hookrightarrow H$ is a map $\phi \triangleq \phi^v \uplus \phi^e \uplus \phi^i \uplus \phi^o$, 
	assigning nodes, edges, inner and outer names with the following constraints:
	\begin{description}
		\item[(L1)\label{def:dlge-1}] $\phi^v : V_G \rightarrowtail V_H$ and $\phi^e : E_G \rightarrowtail E_H$ are
            injective;
		\item[(L2)\label{def:dlge-2}] $ctrl_G = ctrl_H \circ \phi^v$;
		\item[(L3)\label{def:dlge-3}] $\phi^i : Y_H^- \uplus X_H^+ \uplus P_H^+ \rightharpoonup X_G^+ \uplus Y_G^-
            \uplus P_G^+$ is a partial map as follows: \vspace{-1ex}
			\begin{equation*}
				\phi^i(x) \triangleq
				\begin{cases*}
					\phi^{i^-}(x) & if $x \in Y_H^- \uplus P_H^+$ \\
					\phi^{i^+}(x) & if $x \in X_H^+ \uplus P_H^+$
				\end{cases*}
        \quad\text{where}\quad
        \begin{array}{l}
          \phi^{i^-} : Y_H^- \uplus P_H^+ \rightharpoonup Y_G^- \uplus P_G^+ \\ \phi^{i^+} : X_H^+ \uplus P_H^+ \rightharpoonup X_G^+ \uplus P_G^+ \\
          dom(\phi^{i^+}) \cap dom(\phi^{i^-}) = \varnothing
        \end{array}
\vspace{-1ex}
			\end{equation*}
		\item[(L4)\label{def:dlge-4}] $\phi^o : X_G^- \uplus Y_G^+ \rightharpoonup E_H \uplus X_H^- \uplus Y_H^+
            \uplus P_H^-$ is a partial map s.t.:
			\begin{equation*}
				\phi^o(y) \triangleq
				\begin{cases*}
					\phi^{o^-}(y) & if $y \in X_G^-$ \\
					\phi^{o^+}(y) & if $y \in Y_G^+$
				\end{cases*}
        \quad\text{where}\quad
        \begin{array}{l}
        \phi^{o^-} : X_G^- \rightharpoonup E_H \uplus X_H^- \uplus P_H^-
        \\
        \phi^{o^+} : Y_G^+
        			\rightharpoonup E_H \uplus Y_H^+ \uplus P_H^-
        \end{array}
\vspace{-1ex}
			\end{equation*}
		\item[(L5a)\label{def:dlge-5a}] $img(\phi^e) \cap img(\phi^{o}) = \varnothing$;
		\item[(L5b)\label{def:dlge-5b}] $\forall v \in V_G, \forall j \in ar(ctrl(v))~.~\phi^{i}((\phi^v(v), j)) = \bot$;
		\item[(L6a)\label{def:dlge-6a}] $\phi^p \circ link_G^{-1} |_{E_G} = link_H^{-1} \circ \phi^e$;
		\item[(L6b)\label{def:dlge-6b}] $\forall v \in V_G, \forall i \in ar(ctrl(v))~.~\phi^p \circ link_G^{-1}((v, i)) = link_H^{-1}
			\circ \phi^{port}((v,i))$;
		\item[(L7)\label{def:dlge-7}] $\forall p \in dom(\phi^i) : link_H(p) = (\phi^o \uplus \phi^e) (link_G \circ \phi^i(p))$.
	\end{description}
	where $\phi^p \triangleq \phi^{i^+} \uplus \phi^{o^-}\uplus \phi^{port}$ and $\phi^{port} : P_G \rightarrowtail P_H$
	is $\phi^{port}(v,i) \triangleq (\phi^v(v), i)$.
\end{definition}

The first three conditions are on the single sub-maps of the embedding. Conditions \ref{def:dlge-5a}
and~\ref{def:dlge-5b} ensures that no components (except for outer names) are identified; condition \ref{def:dlge-6a}
imposes that points connected by the image of an edge are all covered. Finally, conditions \ref{def:dlge-2}, 
\ref{def:dlge-6b} and \ref{def:dlge-7} ensure that the guest structure is preserved i.e.~node controls and point
linkings are preserved.

\paragraph{Place graph}
Like link graph embeddings, place graph embeddings are just a structure preserving injective map from nodes along with
suitable maps for the inner and outer interfaces. In particular, a site is mapped to the set of sites and nodes that
are ``put under it'' and a root is mapped to the host root or node that is ``put over it'' splitting the host place
graphs in three parts: the guest image, the context and the parameter (which are above and below the guest image).

\begin{definition}[Place graph embedding {\cite[Def~7.5.4]{hojsgaard2012bigraphical}}]\label{def:pge}
	Let $G : n_G \to m_G$ and $H : n_H \to m_H$ be two place graphs. 
	A place graph embedding $\phi : G \emb H$ is a map
	$\phi \defeq \ephi v \uplus \ephi s \uplus \ephi r$
	(assigning nodes, sites and regions respectively)
	subject to the following conditions:
	\begin{description}
	\item[(P1)\label{def:pge-1}]
		$\ephi v : V_G \mono V_H$ is injective;
	\item[(P2)\label{def:pge-2}]
		$\ephi s : n_G \mono \wp(n_H \uplus V_H)$ is fully injective;
	\item[(P3)\label{def:pge-3}]
		$\ephi r : m_G \to V_H \uplus m_H$ in an arbitrary map;
	\item[(P4)\label{def:pge-4}]
		$\rng(\ephi v) \cap \rng(\ephi r) = \emptyset$ and $\rng(\ephi v) \cap \bigcup \rng(\ephi s) = \emptyset$;
	\item[(P5)\label{def:pge-5}]
			$\forall r \in m_G : \forall s \in n_G : \prnt_H^*\circ \ephi r(r) \cap \ephi s(s) = \emptyset$;
	\item[(P6)\label{def:pge-6}]
		$\ephi c \circ \restr{\prnt_G^{-1}}{V_G} = \prnt_H^{-1}\circ \ephi v$;
	\item[(P7)\label{def:pge-7}]
				$\ctrl_G = \ctrl_H \circ \ephi v$;
	\item[(P8)\label{def:pge-8}]
				$\forall c \in n_G \uplus V_G : \forall c' \in \ephi c(c) : 
				(\ephi f \circ \prnt_G)(c) = \prnt_H(c')$;
	\end{description}
	where $\prnt_H^*(c) = \bigcup_{i < \omega} \prnt^i(c)$,
	$\ephi f \defeq \ephi v \uplus \ephi{r}$, and
	$\ephi c \defeq \ephi v \uplus \ephi{s}$.
\end{definition}

These conditions follow the structure of \cref{def:dlge}, the main difference is
\ref{def:pge-5} which states that the image of a root cannot be the descendant of the image of another. Conditions
\ref{def:pge-1}, \ref{def:pge-2} and \ref{def:pge-3} are on the three sub-maps composing the embedding; 
\ref{def:pge-4} and \ref{def:pge-5} ensure that no components are identified; \ref{def:pge-6} imposes surjectivity on children and the last two conditions require the guest structure to be preserved by the embedding.

\paragraph{Directed Bigraph}
Finally, a directed bigraph embedding can be defined as a pair composed by an directed link graph embedding and a place graph embedding, with a consistent interplay of these two structures.
The interplay is captured by two additional conditions ensuring that points (resp. handles) in the image of guest upward (resp. downward) inner names reside in some parameter defined by the place graph embedding (i.e.~descends from the image of a site).

\begin{definition}[Directed bigraph embedding]\label{def:dbge}
	Let $G: \langle n_G, X_G \rangle \to \langle m_G, Y_G \rangle$ and $H: \langle n_H, X_H \rangle \to \langle m_H,
	Y_H \rangle$ be two directed  bigraphs. A directed bigraph embedding is a map $\phi : G \hookrightarrow H$ 
	given by a place graph embedding $\phi^P : G^P \hookrightarrow H^P$ and a link graph embedding 
	$\phi^L : G^L \hookrightarrow H^L$ subject to the following constraints:
	\begin{description}
    \item[(B1)\label{def:dbge-1}] $dom(\phi^{i^+}) \subseteq X_H^+ \uplus \{(v,i) {\in} P_H^+ ~|~ \exists s
        \in n_G, k \in \N : prnt_H^k(v) \in \phi^s(s) \}$;
    \item[(B2)\label{def:dbge-2}] $img(\phi^{o^-}) \subseteq X_H^- \uplus \{(v,i) {\in} P_H^- ~|~ \exists s
        \in n_G, k \in \N : prnt_H^k(v) \in \phi^s(s) \}$.
	\end{description}
\end{definition}

\section{Implementing the embedding problem in CSP}
\label{sec:csp}

In this Section we present a constraint satisfaction problem that models the directed bigraph embedding problem.
The encoding is based solely on integer linear constraints and is proven to be sound and complete.

Initially, we present the encoding for the directed link graph embedding problem and for the place graph embedding problem.
Then we combine them providing some additional ``gluing constraints'' to ensure the consistency of the two sub-problems.
The resulting encodings contains 37 constraint families (reflecting the complexity of the problem definition, see \cref{sec:emb}); hence we take advantage of the orthogonality of link and place structures for the sake of both exposition and adequacy proofs.
We observe that the overall number of variables and constraints produced by the encoding is polynomially bounded with respect to the size of the involved bigraphs, i.e., the number of nodes and edges.

\begin{figure}[t]
	\centering
	\begin{tikzpicture}[>=stealth,scale=.7,font=\small,
		dot/.style={circle,fill=black,minimum size=4pt,inner sep=0, outer sep=3pt},
		var/.style={->},
		lnk/.style={draw=green,thick}]
		\node[dot] (n0) at (0,0) {}; %
		\node[dot] (n1) at (-1,-1) {};
		\node[dot] (n2) at (1,1) {};
		\node[dot] (n3) at (4,0) {}; %
		\node[dot] (n4) at (3,1) {};
		\node[dot] (n5) at (5,-1) {};
		\node[dot] (n6) at (0,-4) {}; %
		\node[dot] (n7) at (-1,-3) {};
		\node[dot] (n8) at (1,-5) {};
		\node[dot] (n9) at (4,-4) {}; %
		\node[dot] (n10) at (3,-5) {};
		\node[dot] (n11) at (5,-3) {};
		
		\draw[var,bend right] (n0) to (n3);
		\draw[var,bend left] (n0) to (n6);
		\draw[var,bend left] (n0) to (n7);
		\draw[var,bend left] (n0) to (n8);
		\draw[var,bend left] (n9) to (n3);
		\draw[var,bend left] (n10) to (n3);
		\draw[var,bend left] (n11) to (n3);
				
		\coordinate (l0) at (2,-5.7);
		\coordinate (l1) at (2,-5.8);
		
		\draw [lnk,out=-90,in=180] (n6) to (l0);		
		\draw [lnk, out=-80,in=180] (n8) to (l0);
		\draw [lnk,<-,out=-120,in=0] (n10) to (l0);
		\draw [lnk,out=-90,in=180] (n7) to (l1);
		\draw [lnk,<-,out=-90,in=0] (n11) to (l1);
		
		\node[] (p0) at (2,2) {Network variables};
		\draw[] (p0.south east) -- (p0.south west);
		\draw[] ($(p0.south east)!.5!(p0.south west)$) -- ++(0,-2);
		
		\node[] (p1) at (2,-6.6) {Guest linking};
		\draw[] (p1.north east) -- (p1.north west);
		\draw[] ($(p1.north east)!.5!(p1.north west)$) -- ++(0,.3);
		
		\node[] (p2) at (-3,1) {Host points};
		\draw[shorten >=6pt] (p2.south west) -- (p2.south east) -- ++ (1,0)-- (n0);
		
		\node[] (p3) at (7.5,1) {Host handles};
		\draw[shorten >=6pt] (p3.south east) -- (p3.south west) -- ++ (-1.3,0)-- (n3);

		\node[] (p4) at (-2.85,-5) {Guest points};
		\draw[shorten >=20pt] (p4.south west) -- (p4.south east) -- ++ (.25,0)-- (n6);
		
		\node[] (p5) at (7.35,-5) {Guest handles};
		\draw[shorten >=20pt] (p5.south east) -- (p5.south west) -- ++ (-.6,0)-- (n9);
	\end{tikzpicture}
	
	\caption{Schema of the multi-flux network encoding.}
	\label{fig:dlge-flux-net}
\end{figure}
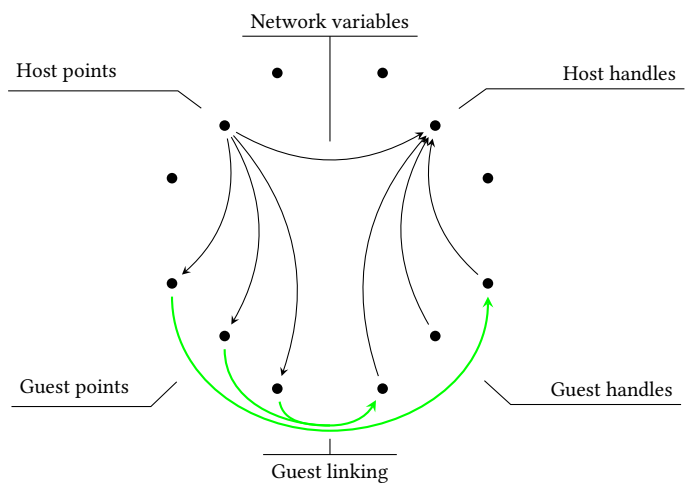

\begin{figure}
\begin{spreadlines}{\cspjot}
	\begin{alignat}{2}
	N_{h,h'} \in \{0,\dots,|\link^{-1}_H(h')|\} &\qquad&& 
		\begin{array}{l}
            h \in E_G \uplus Y_G^{+} \uplus X_G^{-} \uplus P_G^{-}\text{,  } 
            h' \in E_H \uplus Y_H^+ \uplus X_H^{-} \uplus P_H^{-}
    	\end{array}
		\label{eq:dlge-var-1}\\
	N_{p,h'} \in \{0,1\} &\qquad&& 
		\begin{array}{l}
            h' \in E_H \uplus Y_H^{+} \uplus X_H^{-} \uplus P_H^{-}\text{,  } 
            p \in link_H^{-1}(h') 
        \end{array}
		\label{eq:dlge-var-2}\\
	N_{p,p'} \in \{0,1\} &\qquad&& 
		\begin{array}{l}
            p' \in X_G^{+} \uplus P_G^{+} \uplus Y_G^{-}\text{, } 
            p \in X_H^{+} \uplus P_H^{+} \uplus Y_H^{-}
        \end{array}
		\label{eq:dlge-var-3}\\
	F_{h,h'} \in \{0,1\} &\qquad&& 
		\begin{array}{l}
            h \in E_G \uplus Y_G^{+} \uplus X_G^{-} \uplus P_G^{-}\text{,  } 
            h' \in E_H \uplus Y_H^{+} \uplus X_H^{-} \uplus P_H^{-}
        \end{array}
		\label{eq:dlge-var-4}\\
	\sum_{k}N_{p,k} = 1 &\qquad&& 
		\begin{array}{l}
			p \in X_H^{+} \uplus P_H^{+} \uplus Y_H^{-} 
		\end{array}
		\label{eq:dlge-cs-1}\\
	\sum_{k}N_{k,h} = |\link^{-1}_H(h)| &\qquad&& 
		\begin{array}{l}
			h \in E_H \uplus Y_H^{+} \uplus X_H^{-} \uplus P_H^{-}
		\end{array}
		\label{eq:dlge-cs-2}\\
	\sum_{k}N_{h,k} = \sum_{p \in \link_G^{-1}(h)}\sum_{k}N_{k,p} &\qquad&& 
		\begin{array}{l}
        h \in E_G \uplus Y_G^{+} \uplus X_G^{-} \uplus P_G^{-}
		\end{array}
		\label{eq:dlge-cs-3}\\
	\sum_{k}N_{k,p} \leq 1 &\qquad&& 
		\begin{array}{l}
			p \in X_G^{+} \uplus P_G^{+} \uplus Y_G^{-} 
		\end{array}
		\label{eq:dlge-cs-4}\\
	N_{p,p'} = 0 &\qquad&&
		\begin{array}{l}
			p' \in P_G^{+}\text{, }p \in X_H^{+} \uplus Y_H^{-} 
		\end{array}
		\label{eq:dlge-cs-5}\\
	\frac{N_{h,h'}}{|\link_H^{-1}(h')|} \leq F_{h,h'} \leq N_{h,h'} &\qquad&&
		\begin{array}{l}
            h \in E_G \uplus Y_G^{+} \uplus X_G^{-} \uplus P_G^{-}\text{, }
            h' \in E_H \uplus Y_H^{+} \uplus X_H^{-} \uplus P_H^{-}\text{, }\\
            link_G^{-1}(h) \neq \varnothing\text{, }link_H^{-1}(h') \neq \varnothing
		\end{array}
		\label{eq:dlge-cs-6}\\
	N_{p,p'} \leq F_{h,h'} &\qquad&&
		\begin{array}{l}
		    h \in E_G \uplus Y_G^{+} \uplus X_G^{-} \uplus P_G^{-}\text{, }
            h' \in E_H \uplus Y_H^{+} \uplus X_H^{-} \uplus P_H^{-}\text{, }\\
            p \in link_G^{-1}(h)\text{, } p' \in link_H^{-1}(h')
		\end{array}
		\label{eq:dlge-cs-7}\\
	F_{h,h'} \leq \sum_{\substack{p \in \link_G^{-1}(h)\\ p' \in \link_H^{-1}(h')}} N_{p,p'} &\qquad&&
		\begin{array}{l}
            h \in E_G \uplus Y_G^{+} \uplus X_G^{-} \uplus P_G^{-}\text{, }
            h' \in E_H \uplus Y_H^{+} \uplus X_H^{-} \uplus P_H^{-}\text{, }\\
            link_G^{-1}(h) \neq \varnothing\text{, } link_H^{-1}(h') \neq \varnothing
		\end{array}
		\label{eq:dlge-cs-8}\\
	\sum_{k}F_{h,k} = 1 &\qquad&& 
		\begin{array}{l}
			h \in E_G \uplus Y_G^{+} \uplus X_G^{-} \uplus P_G^{-} 
		\end{array}
		\label{eq:dlge-cs-9}\\
	N_{p,h'} + F_{h,h'} \leq 1 &\qquad&&
		\begin{array}{l}
            h \in E_G\text{, }h' \in E_H \uplus Y_H^{+} \uplus X_H^{-} \uplus P_H^{-}\text{, }
            p \in link_H^{-1}(h')
		\end{array}
		\label{eq:dlge-cs-10}\\		
	F_{h,h'} + F_{h'',h'} + F_{h''', h'} + F_{h'^v, h'} \leq 1 &\qquad&&
		\begin{array}{l}
            h \in E_G\text{, } h' \in Y_H^{+} \uplus X_H^{-} \uplus P_H^{-}\text{, }
            h'' \in Y_G^{+}\text{, } h''' \in X_G^{-}\text{, } h'^{v} \in P_G^{-} 
		\end{array}
		\label{eq:dlge-cs-11}\\
	F_{h,h'} = 0 &\qquad&&
		\begin{array}{l}
			h \in E_G\text{, }h' \in Y_H^{+} \uplus X_H^{-} \uplus P_H^{-}
		\end{array}
		\label{eq:dlge-cs-12}\\
	F_{h,h'} \leq 1 &\qquad&&
		\begin{array}{l}
			h \in E_G \uplus Y_G^{+} \uplus X_G^{-} \uplus P_G^{-} \text{, } h' \in E_H 
		\end{array}
		\label{eq:dlge-cs-13}\\
	N_{p,p'} = 0 &\qquad&&
		\begin{array}{l}
            v \in V_G\text{, }v' \in V_H\text{, }
            ctrl_G(v) = ctrl_H(v) = c\text{, }i \neq i' \leq c\text{, }\\
            p = (v,i) \in P_G^{+} \uplus P_G^{-}\text{, }
            p' = (v',i') \in P_H^{+} \uplus P_H^{-}
		\end{array}	
		\label{eq:dlge-cs-14}\\
	N_{p,p'} = 0 &\qquad&&
		\begin{array}{l}
            v \in V_G\text{, }v' \in V_H\text{, }
            ctrl_G(v) \neq ctrl_H(v)\text{, }
            p = (v,i) \in P_G^{+} \uplus P_G^{-}\text{, }\\
            p' = (v',i') \in P_H^{+} \uplus P_H^{-}
		\end{array}	
		\label{eq:dlge-cs-15}\\
	\sum_{j \leq c} N_{(v,j),(v',j)} = c\cdot N_{p,p'} &\qquad&&
		\begin{array}{l}
            v \in V_G\text{, }v' \in V_H\text{, }
            ctrl_G(v) = ctrl_H(v) = c\text{, }i \leq c\text{, }\\
            p = (v,i) \in P_G^{+} \uplus P_G^{-}\text{, }
            p' = (v',i') \in P_H^{+} \uplus P_H^{-}
		\end{array}
		\label{eq:dlge-cs-16}\\
	N_{p,p'} = 0 &\qquad&&
		\begin{array}{l}
			p \in P_H^+\text{, }p' \in X_G^+ \uplus Y_G^-
		\end{array}
		\label{eq:dlge-cs-17}
	\end{alignat}
\end{spreadlines}
\caption{Constraints of \textsc{DLGE}[$G,H$].}
\label{fig:dlge-csp}
\end{figure}

\subsection{Directed Link Graphs}\label{sec:dlge-csp}

Let us fix the guest and host  bigraphs $G : X_G \to Y_G$ and $H : X_H \to Y_H$.
We characterize the embeddings of $G$ into $H$ as the solutions of a suitable \emph{multi-flux problem} which we denote as \textsc{DLGE}[$G,H$].
The main idea is to see the host points (i.e.~positive ports, upward inner names and downward outer names) as sources, and the handles (i.e.~edges, negative ports, upward outer names and downward inner names) as sinks (see \cref{fig:dlge-flux-net}).
Each point outputs a flux unit and each handle inputs one unit for each point it links.
Units flow towards each point handle following $H$ edges and optionally taking a ``detour'' along the linking structure of the guest $G$ (provided that some conditions regarding structure preservation are met). 
The formal definition of the flux problem is in \cref{fig:dlge-csp}.

The flux network reflects the linking structure and contains an edge connecting each point to its handle; these edges have an integer capacity limited to $1$ and are represented by the variables defined in \eqref{eq:dlge-var-2}. 
The remaining edges of the network are organised in two complete biparted graphs: one between guest and host handles andù
èò
one between guest and host points.  Edges of the first sub-network are described by the variables in
\eqref{eq:dlge-var-1} and their capacity is bounded by the number of points linked by the host handle since this is
the maximum acceptable flux and corresponds to the case where each point passes through the same hyper-edge of the
guest link graph.  Edges of the second sub-network are described by the variables in \eqref{eq:dlge-var-3} and, like
the first group of links, have their capacity limited to $1$; to be precise, some of these variables will never assume
a value different from $0$ because guest points can receive flux from anything but the host ports (as expressed by
constraint \eqref{eq:dlge-cs-5}). Edges for the link structure of the guest are presented implicitly in the flux
preservation constraints (see constraint \eqref{eq:dlge-cs-3}). In order to fulfil the injectivity conditions of link
embeddings, some additional \emph{flux variables} (whereas the previous are \emph{network variables}) are defined by
\eqref{eq:dlge-var-4}. These are used to keep track and separate each flux on the bases of the points handle.

The constraint families \eqref{eq:dlge-cs-1} and \eqref{eq:dlge-cs-2} define the outgoing and ingoing flux of host
points and handles respectively.  The former has to send exactly one unit considering every edge they are involved
into and the seconds receive one unit for each of their point regardless if this unit comes from the point directly or
from a handle of the guest. 

The linking structure of the guest graph is encoded by the constraint family \eqref{eq:dlge-cs-3} which states that flux
is preserved while passing through the guest i.e.~the output of each handle has to match the overall input of the
points it connects.

Constraints \eqref{eq:dlge-cs-4}, \eqref{eq:dlge-cs-5}, \eqref{eq:dlge-cs-14}, \eqref{eq:dlge-cs-15},
(\ref{eq:dlge-cs-16}--\ref{eq:dlge-cs-17}) shape the flux in the sub-network linking guest and host points. 
Specifically, \eqref{eq:dlge-cs-4} requires that each point from the guest receives at most one unit; this is needed
when we want to be able to embed a redex where some points (e.g. upward inner names) would not match with an entity of
the agent and (those points) would be deleted anyway when composing the resulting agent back.
Constraints \eqref{eq:dlge-cs-5}, \eqref{eq:dlge-cs-14} and \eqref{eq:dlge-cs-15} disable edges between guest
ports and host inner names, between mismatching ports of matching nodes and between ports of mismatching nodes.
Constraint \eqref{eq:dlge-cs-17} ensures that ascending inner names or descending outer names of the redex are not
matched with positive ports of the agent. Finally, the flux of ports of the same node has to act compactly, as expressed by \eqref{eq:dlge-cs-16}: if there is flux between the $i$-th ports of two nodes, then there should be flux between every other 
ports.

Constraints \eqref{eq:dlge-cs-6}, \eqref{eq:dlge-cs-7} and \eqref{eq:dlge-cs-8} relate flux and network variables
ensuring that the formers assume a true value if, and only, if there is actual flux between the corresponding guest and
host handles. In particular, \eqref{eq:dlge-cs-7} propagates the information about the absence of flux between handles
disabling the sub-network linking handles points and, \emph{vice versa}, \eqref{eq:dlge-cs-8} propagates the
information in the other way disabling flux between handles if there is no flux between their points. 

The remaining constraints prevent fluxes from mixing.
Constraint \eqref{eq:dlge-cs-9} requires guest handles to send their output to exactly one destination thus rendering the sub-network between handles a function assigning guest handles to host handles.
This mapping is subject to some additional conditions when edges are involved: \eqref{eq:dlge-cs-12}
and \eqref{eq:dlge-cs-13} ensure that the edges are injectively mapped to edges only, \eqref{eq:dlge-cs-11} forbids
host outer names to receive flux from an edge and an outer name at the same time. Finally, \eqref{eq:dlge-cs-10} states that the output of host points cannot bypass the guest if there is flux between its handle
and an edge from the guest.

\paragraph{Adequacy}
Let $\vec N$ be a solution of \textsc{DLGE[$G,H$]}. The corresponding link graph embedding $\phi : G \emb H$ is defined as:
\begin{alignat*}{3}
    \phi^v(v) & \triangleq v' \in V_H  \text{ if }   \exists i : N_{(v,i), (v',i)} = 1 
    &\qquad
    \phi^e(e) & \triangleq e' \in E_H  \text{ if }  F_{e,e'} = 1 \\
    \phi^i(x) & \triangleq
\begin{cases*}
\phi^{i^-}(x) & if $x \in Y_H^- \uplus P_H^+$ 
\\
\phi^{i^+}(x) & if $x \in X_H^+ \uplus P_H^+$
\end{cases*}
    &
    \phi^o(y) & \triangleq
\begin{cases*}
\phi^{o^-}(y) & if $y \in X_G^-$ \\
\phi^{o^+}(y) & if $y \in Y_G^+$
\end{cases*}
\end{alignat*}
where $\phi^{o^-}(y) \triangleq y' \in X_H^- \uplus P_H^- \text{ if } F_{y,y'} = 1$, $\phi^{o^+}(y) \triangleq y' \in Y_H^+ \uplus P_H^-  \text{ if }  F_{y,y'} = 1$, $\phi^{i^-}(x) \triangleq x' \in Y_G^- \uplus P_G^+  \text{ if }  N_{x,x'} = 1$, $\phi^{i^+}(x) \triangleq x' \in X_G^+ \uplus P_G^+  \text{ if }  N_{x,x'} = 1$, and $dom(\phi^{i^+}) \cap dom(\phi^{i^-}) = \varnothing$.
It is easy to check that these components of $\phi$ are well-defined and compliant with \cref{def:dlge}.

On the other way around,
let $\phi : G \hookrightarrow H$ be a link graph embedding. The corresponding solution $\vec N$ of $DLGE[G,H]$
is defined as follows:
\begin{align*}
    N_{p,p'} & \triangleq
        \begin{cases*}
            1 & if $p \in X_H^+ \uplus Y_H^- \land p' = \phi^i(p)$ \\
            1 & if $p' = (v,i) \in P_G^+ \land p = (\phi^v(v), i)$ \\
            0 & otherwise
        \end{cases*}
    &\quad
    N_{h,h'} & \triangleq
\begin{cases*}
1 & if $h' \in E_H \land h \in E_G \land h' = \phi^e(h)$ \\
1 & if $h' \in Y_H^+ \uplus X_H^- \land h \in Y_G^+ \uplus X_G^- \land h' = \phi^o(h)$ \\
1 & if $h = (v,i) \in P_G^- \land h' = (\phi^v(v), i)$ \\
0 & otherwise
\end{cases*}
    \\
    N_{p,h'} & \triangleq
\begin{cases*}
1 & if $h' = link_H(p) \land \nexists p' : N_{p,p'} = 1$ \\
0 & otherwise
\end{cases*}
        &
    F_{h,h'}& \triangleq
        \begin{cases*}
          0 & if $N_{h,h'} = 0$ \\
          1 & otherwise
        \end{cases*}
\end{align*}
Every constraint of $DLGE[G,H]$ is satisfied by the solution just defined.

The constraint satisfaction problem in \cref{fig:dlge-csp} is sound and complete with respect to the directed link graph embedding problem given in \cref{def:dlge}.

\begin{proposition}[Adequacy of \textsc{DLGE}]
\label{prop:dlge-adequacy}
    For any two concrete directed link graphs $G$ and $H$, there is a bijective correspondence between the directed
    link graph embeddings of $G$ into $H$ and the solutions of \textsc{DLGE[$G,H$]}.

\end{proposition}

\subsection{Place Graphs}

\begin{figure}[t]
	\begin{spreadlines}{\cspjot}
	\begin{alignat}{2}
		M_{h,g} \in \{0,1\} &\qquad&& 
			\begin{array}{l}
				g \in n_G \uplus V_G \uplus m_G\text{, }
				h \in n_H \uplus V_H \uplus m_H
			\end{array}
			\label{eq:pge-var-1}\\
		M_{h,g} = 0 &\qquad&&
			\begin{array}{l}
				g \in n_G \uplus V_G\text{, }h \in m_H
			\end{array}
			\label{eq:pge-cs-1}\\
		M_{h,g} = 0 &\qquad&&
			\begin{array}{l}
				g \in V_G \uplus m_G\text{, }h \in n_H
			\end{array}
			\label{eq:pge-cs-2}\\
		M_{h,g} = 0 &\qquad&&
			\begin{array}{l}
				g \in V_G \text{, }h \in V_H\text{, }
				\ctrl_G(g) \neq \ctrl_H(h)
			\end{array}
			\label{eq:pge-cs-3}\\
		M_{h,g} = 0 &\qquad&&
			\begin{array}{l}
				g \in m_G \text{, }h \notin m_H\text{, }
				v \in \prnt_H^*(h) \cap V_G\text{, }
				\ctrl_G(v) \notin \Sigma_a
			\end{array}
			\label{eq:pge-cs-4}\\
		M_{h,g} \leq M_{h',g'} &\qquad&&
			\begin{array}{l}
				g \notin m_G\text{, }g' \in \prnt_G(g)\text{, }
				h \notin m_H\text{, }h' \in \prnt_H(h)
			\end{array}
			\label{eq:pge-cs-5}\\
		\sum_{h \in V_H \uplus m_H} M_{h,g} = 1 &\qquad&& 
			\begin{array}{l}
				g \in m_G
			\end{array}
			\label{eq:pge-cs-6}\\
		\sum_{h \in n_H \uplus V_H} M_{h,g} = 1 &\qquad&& 
			\begin{array}{l}
				g \in V_G
			\end{array}
			\label{eq:pge-cs-7}\\
		m_G \cdot \sum_{g \in n_G\uplus V_G} M_{h,g} + \sum_{g \in m_G} M_{h,g}
		\leq m_G &\qquad&&
			\begin{array}{l}
				h \in V_H
			\end{array}
			\label{eq:pge-cs-8}\\
		|\prnt_H^{-1}(h)|\cdot M_{h,g} \leq 
			\sum_{\substack{h' \in \prnt_H^{-1}(h),\\ g' \in \prnt_G^{-1}(g)}} 
			M_{h'\!,g'} &\qquad&&
			\begin{array}{l}
				g \in V_G\text{, }h \in V_H
			\end{array}
			\label{eq:pge-cs-9}\\
		|\prnt_G^{-1}(g)\setminus n_G|\cdot M_{h,g} \leq 
			\sum_{\substack{h' \in \prnt_H^{-1}(h)\setminus n_h,\\ g' \in \prnt_G^{-1}(g) \setminus n_g}} 
			M_{h'\!,g'} &\qquad&&
			\begin{array}{l}
				g \in m_G\text{, }h \in V_H
			\end{array}
			\label{eq:pge-cs-10}\\
		M_{h,g} + \sum_{\substack{h' \in \prnt_H^*(h), g'\in m_G}}M_{h',g'} \leq 1 &\qquad&&
			\begin{array}{l}
				g \in V_G\text{, }h\in V_H
			\end{array}
			\label{eq:pge-cs-11}
	\end{alignat}
	\end{spreadlines}
	\caption{Constraints of \textsc{PGE}[$G,H$].}
	\label{fig:pge-csp}
\end{figure}

Let us fix the guest and host place graphs:
$G : n_G \to m_G$ and $H : n_H \to m_H$. We characterize
the embeddings of $G$ into $H$ as the solutions of the
constraint satisfaction problem in \cref{fig:pge-csp}.
The problem is a direct encoding of \cref{def:pge}
as a matching problem presented, as usual, as a bipartite graph.
Sites, nodes and roots of the two place graphs are represented as nodes
and parted into the guest and the host ones. For convenience of exposition, 
graph is complete.

Edges are modelled by the boolean variables defined in \eqref{eq:pge-var-1};
these are the only variables used by the problem. So far, a solution is
nothing more than a relation between the components of guest and host
containing only those pairs connected by an edge assigned a non-zero value.
To capture exactly those assignments that are actual place graph embeddings
some conditions have to be imposed.

Constraints \eqref{eq:pge-cs-1} and \eqref{eq:pge-cs-2}
prevent roots and sites from the host to be matched with nodes or sites
and nodes or roots respectively. \eqref{eq:pge-cs-3} 
disables matching between nodes decorated with different controls.
Constraint \eqref{eq:pge-cs-4} prevents any matching for host nodes
under a passive context (i.e.~have an ancestor labelled with a passive control).
\eqref{eq:pge-cs-5} propagates the matching along the parent map from children
to parents. Constraints \eqref{eq:pge-cs-6} and \eqref{eq:pge-cs-7}
ensure that the matching is a function when restricted to guest nodes and roots
(the codomain restriction follows by \eqref{eq:pge-cs-1} and \eqref{eq:pge-cs-2}).
\eqref{eq:pge-cs-8} says that if a node from the host cannot be
matched with a root or a node/site from the guest at the same time;
moreover, if the host node is matched with a node then it cannot be matched
to anything else.

The remaining constraints are the counterpart of \eqref{eq:pge-cs-5}
and propagate matchings from parents to children.
\eqref{eq:pge-cs-9} applies on matchings between nodes and says
that if parents are matched, then children from the host node are covered
by children from the guest node. In particular, the matching is a perfect assignment
when restricted to guest children that are nodes  (because of \eqref{eq:pge-cs-8})
and is a surjection on those that are sites.
\eqref{eq:pge-cs-10} imposes a similar condition on matchings between
guest roots and host nodes. Specifically, it says that the matching has
to cover child nodes from the guest (moreover, it is injective on them)
leaving child sites to match whatever remains ranging from nothing to all
unmatched children. Finally, \eqref{eq:pge-cs-11} prevent matching from happening inside a parameter.

\paragraph{Adequacy}
Let $\vec M$ be a solution of \textsc{PGE[$G,H$]}. The corresponding
place graph embedding $\phi : G \emb H$ is defined as:
\begin{equation*}
	\ephi{v}(g) \defeq 
		h  \in V_H \text{ if } \exists i : M_{h,g} = 1 \qquad
	\ephi{s}(g) \defeq 
		\{h \in n_h \uplus V_H \mid M_{h,g} = 1\} \qquad
	\ephi{r}(g) \defeq 
		h \in m_H \uplus V_H \text{ if } M_{h,g} = 1
\end{equation*}
These components of $\phi$ are well-defined and
compliant with \cref{def:pge}.
On the opposite direction, let ${\phi : G \emb H}$ be a place
graph embedding. The corresponding solution $\vec M$ 
of \textsc{PGE[$G,H$]} is defined as aside.
It is easy to check that every constraint of \textsc{PGE[$G,H$]}
is satisfied by this solution.
\[M_{h,g} \defeq
	\begin{cases}
		1 & \mbox{if } g \in V_G \land h = \ephi{v}(g) \\
		1 & \mbox{if } g \in m_G \land h = \ephi{r}(g) \\
		1 & \mbox{if } g \in n_G \land h \in \ephi{s}(g) \\
		0 & \mbox{otherwise}
	\end{cases}\]
Hence, the constraint satisfaction problem in \cref{fig:pge-csp}
is sound and complete with respect to the place graph embedding problem
(\cref{def:pge}).
\begin{proposition}[Adequacy of \textsc{PGE}]
\label{prop:pge-adequacy}
For any two concrete place graphs $G$ and $H$,
there is a bijective correspondence between
the place graph embeddings of $G$ into $H$ and
the solutions of \textsc{PGE[$G,H$]}.
\end{proposition}

\subsection{Bigraphs}
Let $G : \face{n_G,X_G} \to \face{m_G,Y_G}$ and $H : \face{n_H,X_H} \to \face{m_H,Y_H}$ be two bigraphs. By taking advantage of the orthogonality of the link and place structures we can define the constraint satisfaction problem capturing bigraph embeddings by simply composing the constraints given above for the link and place graph embeddings and by adding four consistency constraints to relate the solutions of the two problems.
These additional constraint families are reported in \cref{fig:dbge-csp}. The families \eqref{eq:dbge-cs-1} and
\eqref{eq:dbge-cs-2} ensure that solutions for \textsc{DLGE[$G,H$]} and \textsc{PGE[$G,H$]} agree on nodes since the
map $\ephi{v}$ has to be shared by the corresponding link and place embeddings. The families \eqref{eq:dbge-cs-3} and
\eqref{eq:dbge-cs-4} respectively, ensure that positive ports (negative ports resp.) are in the same image as
upward inner names (downward inner names resp.) only if their node is part of the parameter i.e.~only if it is
matched to a site from the guest or it descends from a node that is so. 

\begin{figure}[t]
	\begin{spreadlines}{\cspjot}
	\begin{alignat}{2}
	M_{v,v'} = N_{p,p'} &\qquad&&
        \begin{array}{l}   
            v \in V_H\text{, }v' \in V_G\text{, }
            p = (v,k) \in P_H^+\text{, }p' = (v',k) \in P_G^+
        \end{array}
		\label{eq:dbge-cs-1}\\
	M_{v,v'} = F_{h,h'} &\qquad&&
		\begin{array}{l}   
			v \in V_H\text{, }v' \in V_G\text{, }
			h \in P_G^-\text{, }h' \in P_H^-
		\end{array}
		\label{eq:dbge-cs-2}\\
	\sum_{p' \in X_G^+} N_{p,p'} \leq \sum_{\substack{h \in \prnt_H^*(v), g \in n_G}} M_{h,g} &\qquad&&
		\begin{array}{l}   
			v \in V_H\text{, }p = (v,k) \in P_H^+
		\end{array}
		\label{eq:dbge-cs-3}\\
	\sum_{h \in X_G^-} F_{h,h'} \leq \sum_{\substack{h \in \prnt_H^*(v), g \in n_G}} M_{h,g} &\qquad&&
		\begin{array}{l}   
			v \in V_H\text{, }h' = (v,k) \in P_H^-
		\end{array}
		\label{eq:dbge-cs-4}
	\end{alignat}
	\end{spreadlines}
	\caption{Constraints of \textsc{DBGE}[$G,H$].}
	\label{fig:dbge-csp}
\end{figure}

Conditions \eqref{eq:dbge-cs-3} and \eqref{eq:dbge-cs-4} correspond exactly to \ref{def:dbge-1} and \ref{def:dbge-2}. It thus follow from \cref{prop:dlge-adequacy,prop:pge-adequacy} that the CSP defined by \cref{fig:dlge-csp,fig:pge-csp,fig:dbge-csp} is sound and complete with respect to the bigraph embedding problem given in \cref{def:dbge}.

\begin{theorem}[Adequacy of \textsc{BGE}]\label{prop:bge-adequacy}
For any two concrete bigraphs $G$ and $H$, there is a bijective correspondence between the bigraph embeddings of $G$
into $H$ and the solutions of \textsc{DBGE[$G,H$]}.
\end{theorem}

\section{Experimental results}
\label{sec:experiments}

The reduction algorithm presented in the previous section has been successfully integrated into \libbig, an extensible Java library for manipulating bigraphs and bigraphical reactive systems which can be used for implementing a wide range of tools and it can be adapted to support several extensions of bigraphs \cite{jlibbig}.

The proposed algorithm is implemented by extending the data structures and the models for pure bigraphs to  suit our definition of directed bigraphs.

\begin{figure}[t]
    \begin{subfigure}{\textwidth}
        \centering
        \includegraphics[width=.7\textwidth]{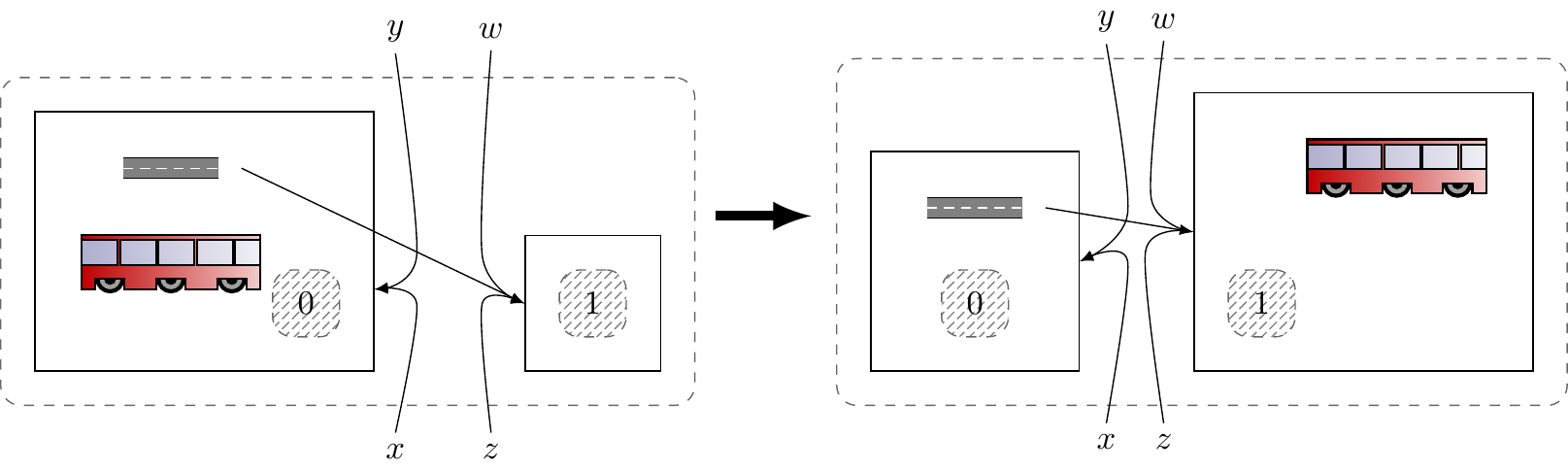}
        \caption{Rewriting rule}
        \label{fig:rew_rule}
        \bigskip
    \end{subfigure}
%

    \begin{subfigure}{\textwidth}
      \centering
      \includegraphics[width=0.7\textwidth]{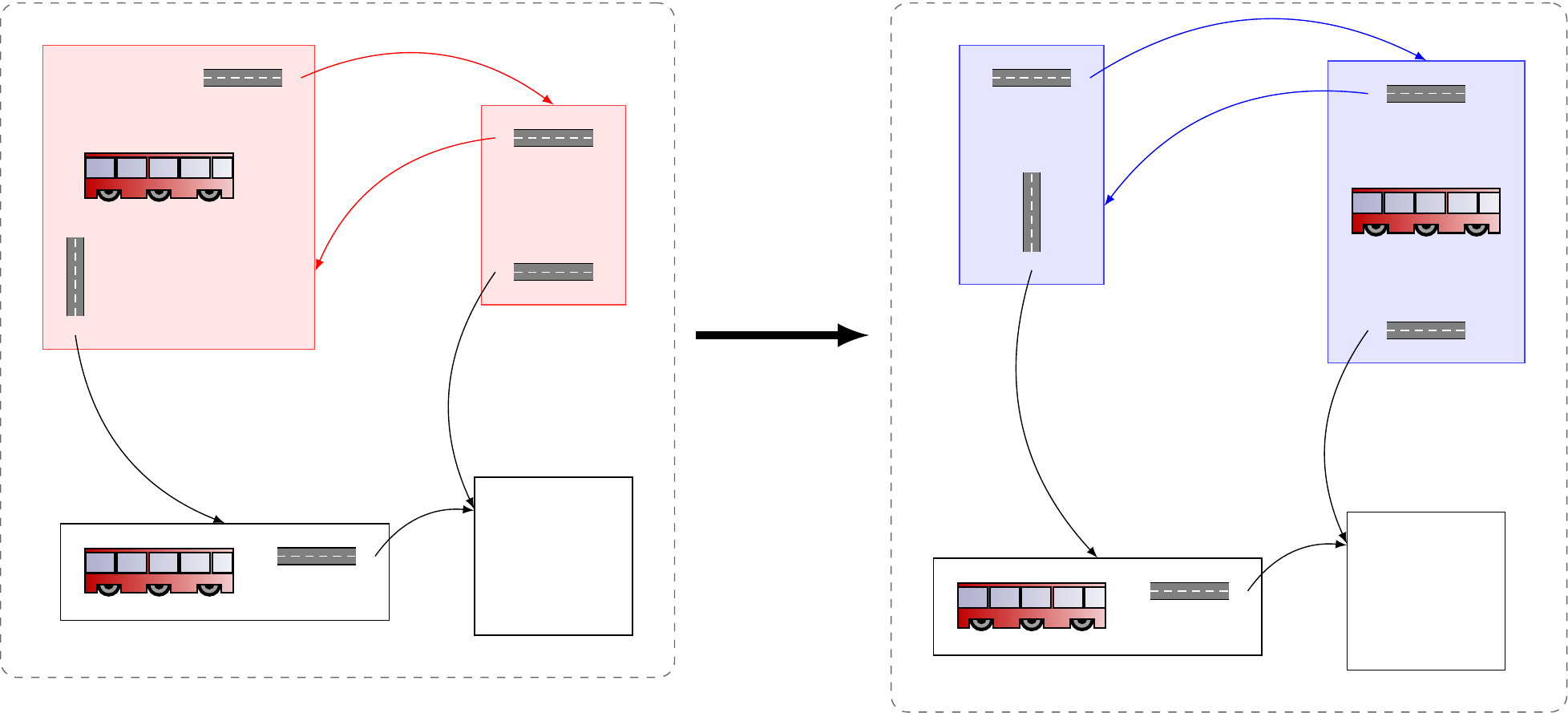}
      \\[2ex]
      \includegraphics[width=0.7\textwidth]{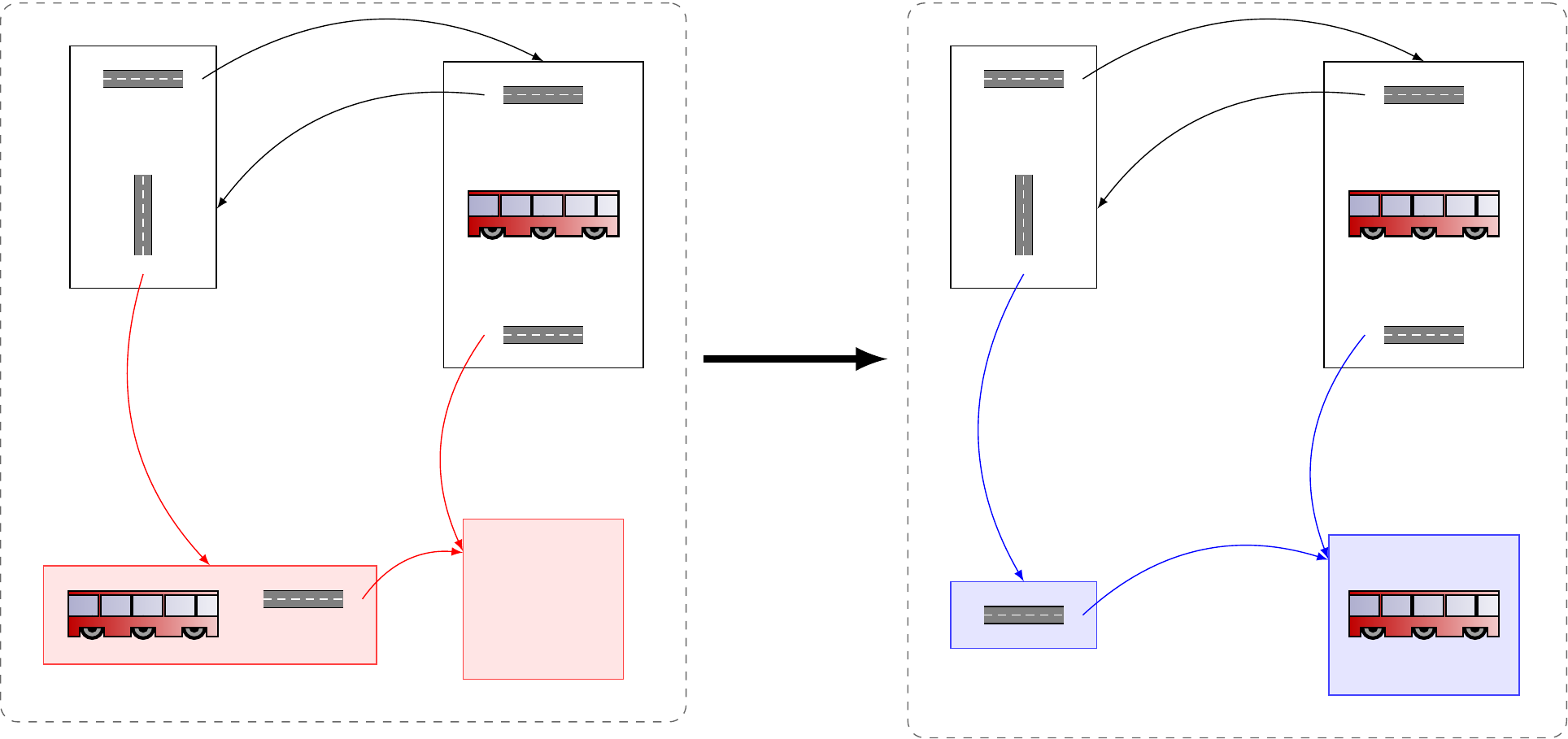}
      \caption{Example of transitions; embeddings of the redex and reactum highlighted in red and blue, respectively.}
      \label{fig:exec1}
    \end{subfigure}

    \caption{Rewriting rule (a) and examples of its application (b).}
    \label{fig:vehicles_exmp}
\end{figure}

In this section we test our implementation by simulating a system in which we want to track the position and the movements of a fleet of vehicles inside a territory divided in ``zones'', which are accessible via ``roads''.
The rewriting rule in question and an example agent can be found in \cref{fig:vehicles_exmp}.

We evaluate the running time of the different components of our algorithm: model construction, CSP resolution, building of the actual embedding and execution of the rewriting rule.
Moreover, we want to evaluate how these performances scale while increasing the size of the agent. 
The parameters used to build the tests are: number of zones, number of cars and ``connectivity degree''. 
The last parameter is a number between 1 and 100 representing the probability of the existence of a connection between two nodes; a value of 100 means that every node is connected to all its neighbours.
All tests have been performed on an Intel Core i7-4710HQ  (4 cores at 3.5GHz), 8 GB of RAM running on ArchLinux with kernel 5.5.2 and using OpenJDK 12.

We consider the following kinds of tests:
\begin{enumerate}
    \item varying number of cars, with fixed number of zones and connectivity degree;
    \item varying number of zones, with fixed number of cars and connectivity degree;
    \item varying connectivity degree, with fixed numbers of zones and cars.
\end{enumerate}
Each test case is made up of four groups of instances, where for each group we choose an increasing value for their fixed parameters.
For each group we choose ten values for its variable parameter.

The instances generation works as follows: for each test case and for each group of that particular test case we generate ten random instances for each combination of the values of the fixed parameters and the variable one.  We then take the average of the running times of those ten random instances.
At the end of the process, for each group we have tested 100 instances, 10 for each value of the variable parameter, so 400 instances for each test case and 1200 in total.

We briefly review the results obtained from these tests.
\def\tabfig#1{%
  \begin{minipage}{.38\textwidth}
    \centering
    \csvautobooktabularcenter{#1.csv}
  \end{minipage}\hfill\begin{minipage}{.60\textwidth}
    \includegraphics[width=\textwidth]{#1.pdf}
  \end{minipage}
}
\begin{figure}[p]
    \tabfig{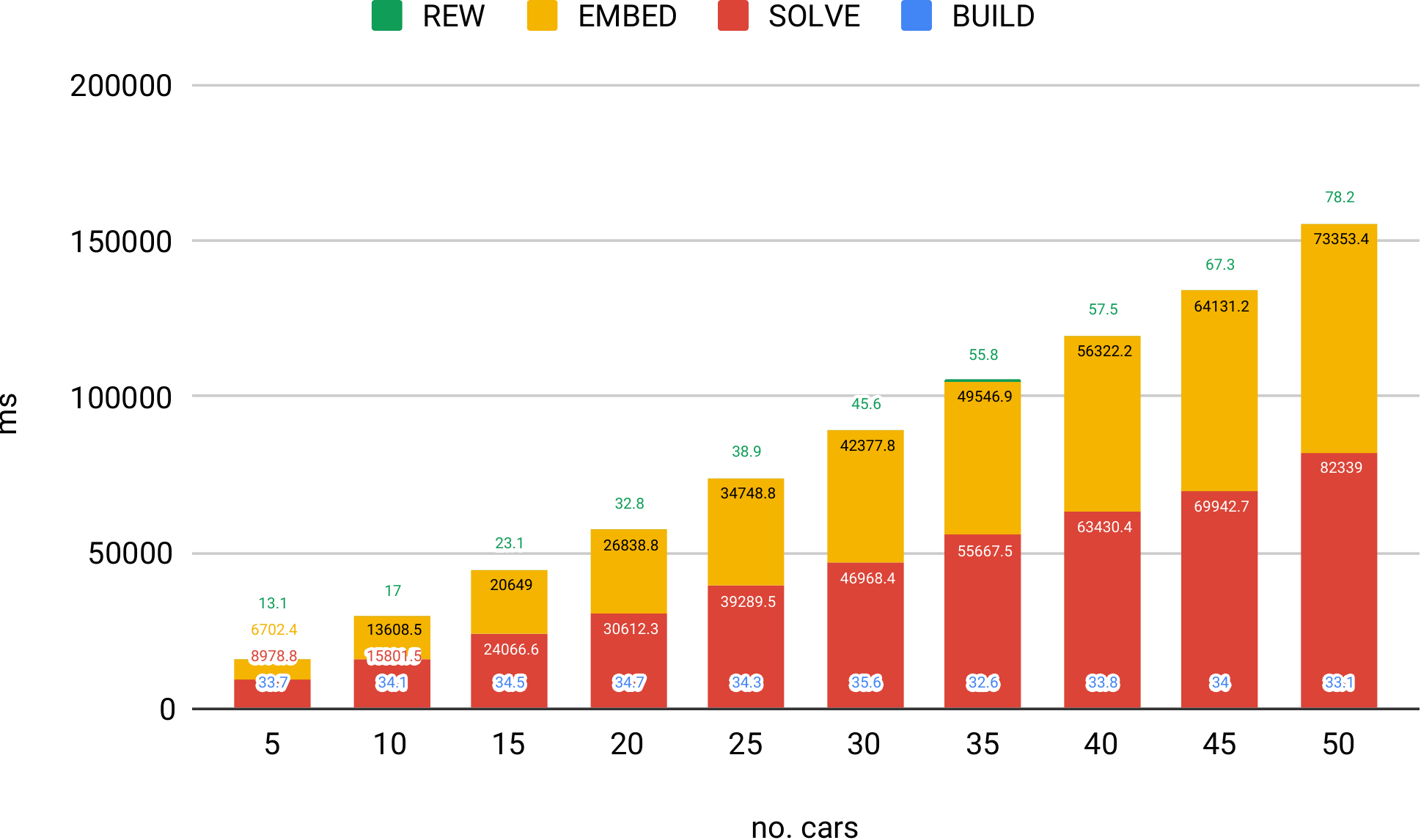}
    \caption{Execution times vs. number of cars, on 11x11 zone grid with 100\% connectivity.}
    \label{fig:times_cars}
\end{figure}
\begin{figure}[p]
    \tabfig{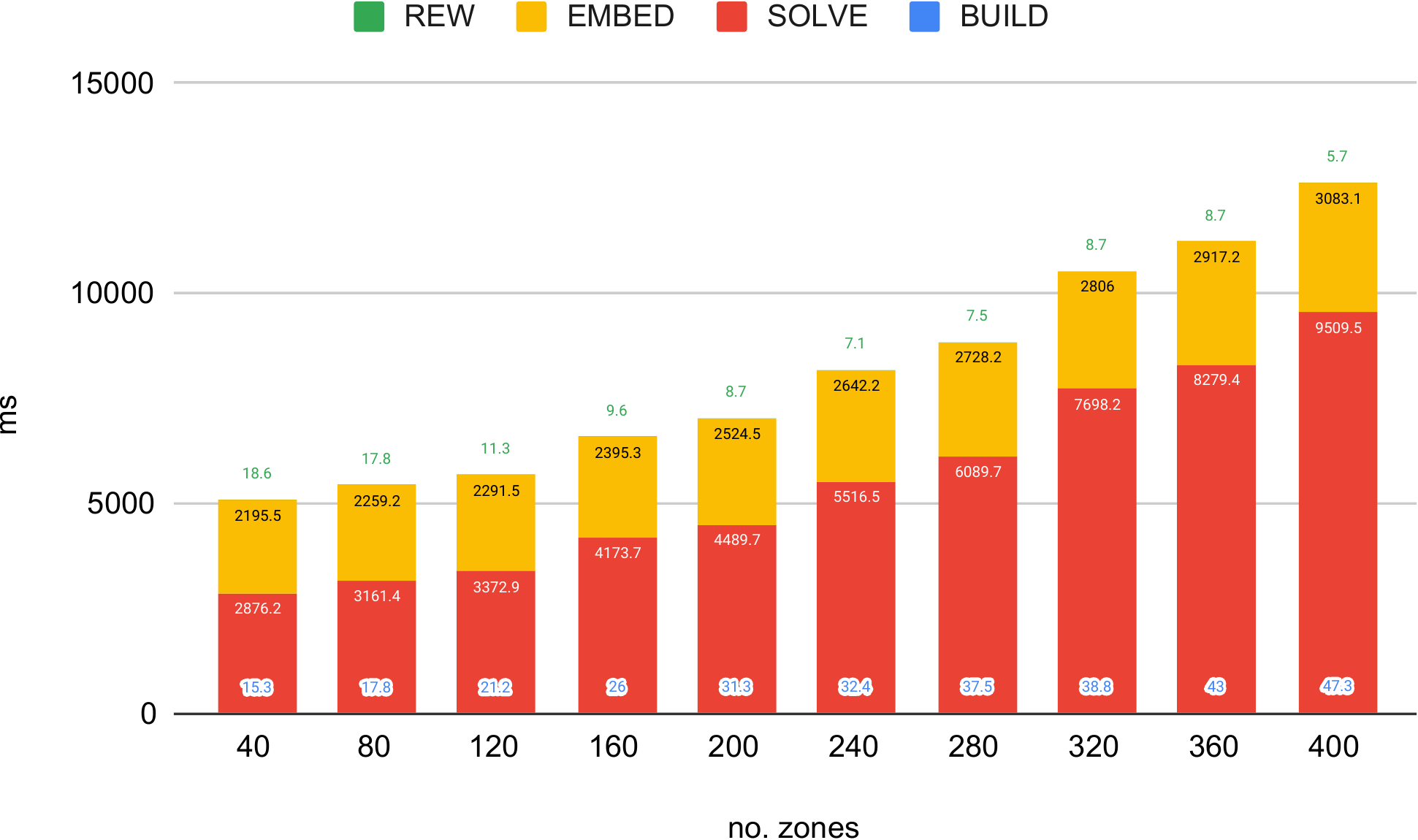}
    \caption{Execution times vs. number of zones, with 70 cars and 100\% connectivity.}
    \label{fig:times_zones}
\end{figure}
\begin{figure}[p]
    \tabfig{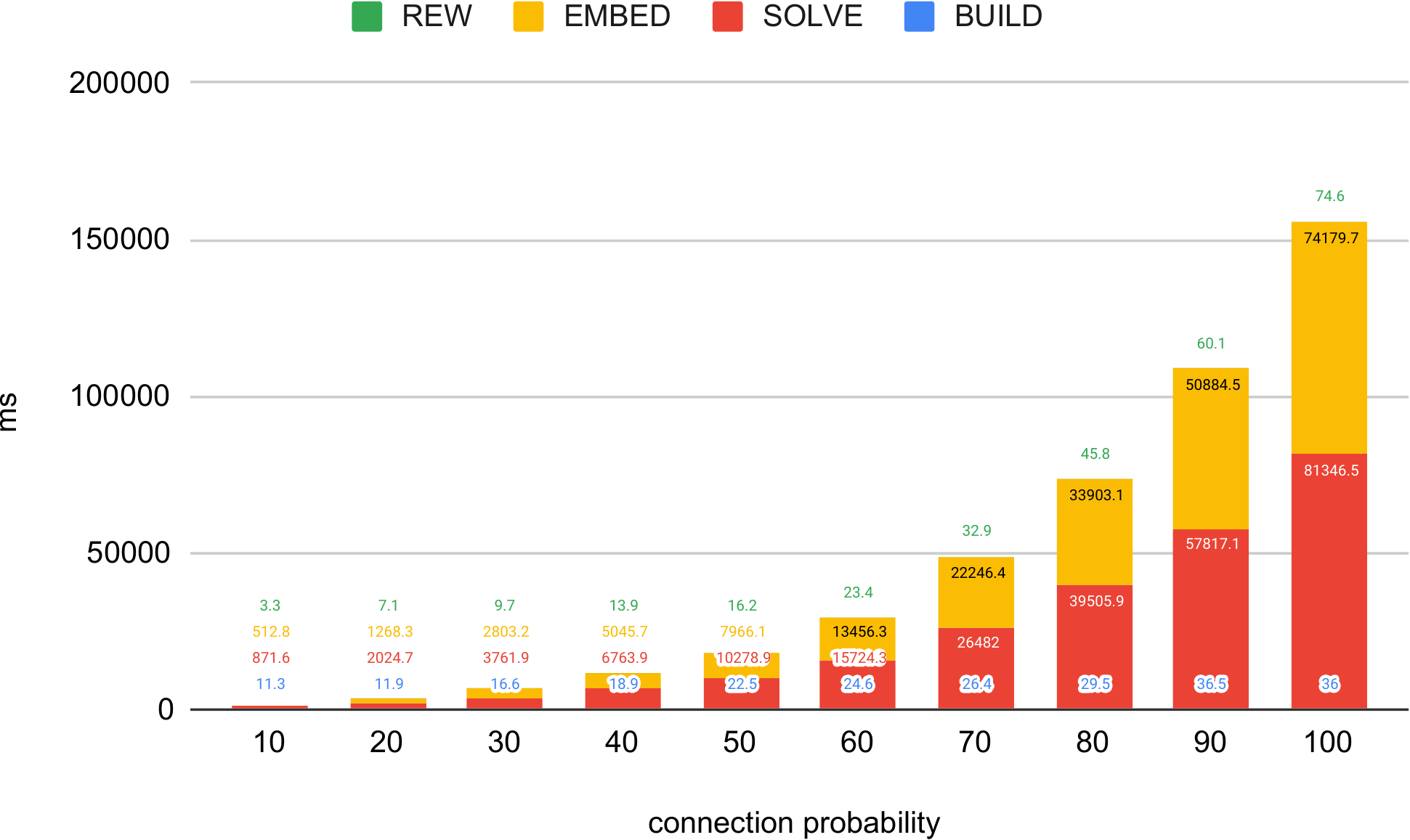}
    \caption{Execution times vs. connectivity, with a 11x11 zone grid and 70 cars.}
    \label{fig:times_density}
\end{figure}

\paragraph{Time vs. Number of Cars}
We evaluate how our implementation scales with an increasing number of cars; see \cref{fig:times_cars}.
We can observe that the (total) execution time increases linearly.
We can also see that the two phases that contribute the most to the total running time are the solving phase of the CSP and the phase in which the embedding is built from the solution of the CSP.
On the other hand, the building and rewriting phases are executed nearly instantly. We observe
that the time needed to build the CSP is almost constant, while the rewriting time scales almost linearly as well.

\paragraph{Time vs. Number of Zones}
We evaluate how our implementation scales with an increasing number of zones; see \cref{fig:times_zones}.
We can see that the running time grows exponentially, especially the resolution time. 
Similarly to the previous test case, the time spent building the CSP and applying the reaction rule is negligible even though we can see that the time necessary to build the CSP increases linearly with the grid size.
We can also observe that there is no correlation between the rewriting time and the number of zones.

\paragraph{Time vs. Connectivity degree}
We evaluate how our implementation scales with an increasing connectivity degree; see \cref{fig:times_density}.
We can see that the running time scales exponentially, no matter the grid size or the number of
cars.
Once again, we see that although increasing, the time spent building the model and applying the rewriting rule is negligible.

\FloatBarrier
\section{Conclusions and future works}\label{sec:concl}
In this paper, we have presented a new version of \emph{directed} bigraphs and bigraphic reactive systems, which subsume previous versions (such as Milner's bigraphs). 
For this kind of bigraphs we have provided a sound and complete algorithm for solving the embedding problem, based on a constraint satisfaction problem.
The resulting model is compact and the a number of variables and linear constraints are polynomially bounded by the size of the guest and host bigraphs.
Differently from existing solutions, this algorithm applies also to non-ground hosts.

The proposed approach offers great flexibility: it can be easily applied also to other extensions of bigraphs and directed bigraphs e.g.~bigraphs with sharing \cite{calder2012process} or local directed bigraphs~\cite{burco2019towards,peressotti2012}.

The algorithm has been successfully integrated into jLibBig, an extensible library for manipulating bigraphical reactive systems. 
The empirical evaluation of the implementation of our algorithm in \libbig\ looks promising.
It cannot be considered a rigorous experimental validation yet, mainly because performance depends on the implementation and the solver and the model is not optimized for any specific solver. 
Moreover, up to now there are no ``official'' (or ``widely recognized'') benchmarks, nor any other algorithms or available tools that solve the directed bigraph embedding problem, to compare with.

An interesting direction for future work would be to extend the algorithm also to stochastic and probabilistic bigraphs \cite{krivine2008stochastic}; this would offer useful modelling and verification tools for quantitative aspects, e.g.~for systems biology \cite{bacci2009framework,damgaard2012formal}.
Approximated and weighted embeddings are supported in \libbig, but still as experimental feature. In fact, the theoretical foundations of these extensions have not been fully investigated yet, suggesting another line of research.

\begin{anonsuppress}
\begin{acks}
	This work was partially supported by the Independent Research Fund Denmark,
	Natural Sciences, grant DFF-7014-00041, 
	and Italian PRIN 2017 grant \emph{IT MATTERS}.
\end{acks}
\end{anonsuppress}

\FloatBarrier

\bibliography{biblio}

\end{document}